\documentclass[onecolumn,preprint,superscriptaddress,nofootinbib,10pt]{revtex4-1}

\usepackage[perpage]{footmisc}
\usepackage{epstopdf}
\usepackage[dvipsnames]{xcolor}
\usepackage{epsfig}
\usepackage{color}
\usepackage{hyperref}
\usepackage{mathbbol}
\usepackage{bookmark}
\usepackage{amsmath}
\usepackage{amssymb}
\usepackage{slashed}
\usepackage{nicefrac}
\usepackage{tikz,pgfplots}
\tikzset{mynode/.style={inner sep=2pt,fill,outer sep=0,circle}}
\usetikzlibrary{external}
\usepgfplotslibrary{patchplots}
\pgfplotsset{compat=1.10}
\pgfplotsset{
  cuslegR/.style={legend image code/.code={
\draw[red,dash pattern=on 1.5pt off 1.5pt on 1.5pt off 1.5pt] (0cm,0cm)     -- (0.5cm,0cm);
\draw[red,solid]  (0.5cm, 0cm) -- (1.cm, 0cm);
}}}
\pgfplotsset{
  cuslegG/.style={legend image code/.code={
\draw[green,dash pattern=on 1.5pt off 1.5pt on 1.5pt off 1.5pt] (0cm,0cm)     -- (0.5cm,0cm);
\draw[green,solid]  (0.5cm, 0cm) -- (1.cm, 0cm);
}}}
\pgfplotsset{
  cuslegB/.style={legend image code/.code={
\draw[blue,dash pattern=on 1.5pt off 1.5pt on 1.5pt off 1.5pt] (0cm,0cm)     -- (0.5cm,0cm);
\draw[blue,solid]  (0.5cm, 0cm) -- (1.cm, 0cm);
}}}
\pgfplotsset{
  cuslegBl/.style={legend image code/.code={
\draw[black,dash pattern=on 1.5pt off 1.5pt on 1.5pt off 1.5pt] (0cm,0cm)     -- (0.5cm,0cm);
\draw[black,solid]  (0.5cm, 0cm) -- (1.cm, 0cm);
}}}
\usepgfplotslibrary{fillbetween}
\usetikzlibrary{positioning}
\usetikzlibrary{calc}
\usetikzlibrary{arrows,shapes,backgrounds}
\pgfdeclarepatternformonly{north east lines wide}%
   {\pgfqpoint{-1pt}{-1pt}}%
   {\pgfqpoint{10pt}{10pt}}%
   {\pgfqpoint{9pt}{9pt}}%
   {
     \pgfsetlinewidth{1.4pt}
     \pgfpathmoveto{\pgfqpoint{0pt}{0pt}}
     \pgfpathlineto{\pgfqpoint{9.1pt}{9.1pt}}
     \pgfusepath{stroke}
    }
\tikzset{
  pics/carc/.style args={#1:#2:#3}{
    code={
      \draw[pic actions] (#1:#3) arc(#1:#2:#3);
    }
  }
}
\usetikzlibrary{decorations.pathreplacing}
\usetikzlibrary{decorations.pathmorphing,patterns}
\usetikzlibrary{decorations.markings,calc}
\definecolor{blue}{HTML}{4169E1}
\definecolor{red}{HTML}{DC143C}
\definecolor{green}{HTML}{2E8B57}
\definecolor{black}{HTML}{000000}
\definecolor{g1}{HTML}{A9A9A9}
\definecolor{g2}{HTML}{696969}
\definecolor{g3}{HTML}{7F7F7F}
\definecolor{g4}{HTML}{D3D3D3}
\newcommand{\he}{$\alpha$}
\newcommand{\hes}{${}^3$He}
\newcommand{\tr}{${}^3$H}
\newcommand{\ls}{\ve{L}\cdot\ve{S}}
\newcommand{\eps}{\epsilon}
\newcommand{\as}{a_s}
\newcommand{\at}{a_t}
\newcommand{\ecm}{E_\textrm{\small c.m.}}

\newcommand{\eg}{\textit{e.g.}\;}
\newcommand{\ie}{\textit{i.e.}\;}
\newcommand{\cf}{\textit{c.f.}\;}
\newcommand{\be}{\begin{equation}}
\newcommand{\ee}{\end{equation}}
\newcommand{\la}{\label}
\newcommand{\ber}{\begin{eqnarray}}
\newcommand{\eer}{\end{eqnarray}}

\newcommand{\bea}{\begin{eqnarray}}
\newcommand{\eea}{\end{eqnarray}}
\newcommand{\beq}{\begin{align}}
\newcommand{\eeq}{\end{align}}

\newcommand{\bt}{B_{^{3}\text{H}}}
\newcommand{\bh}{B_{^{3}\text{He}}}
\newcommand{\bd}{B_\text{D}}
\newcommand{\ba}{B_\alpha}
\newcommand{\rgm}{$\mathbb{R}$GM}

\newcommand{\lam}[1]{$\Lambda=#1~$fm$^{-1}$}
\newcommand{\parg}[1] {\paragraph*{-\,\textit{#1}\,-}}

\newcommand{\ve}[1]{\ensuremath{\boldsymbol{#1}}}

\newcommand{\eftnopi}{\mbox{EFT($\slashed{\pi}$)}}

\begin{document}

\title{Asymmetric regularization of the ground and excited state\\
of the ${}^4$He nucleus}

\author{Johannes~Kirscher}
\email{jkirscher@ccny.cuny.edu}
\affiliation{Department of Physics, The City College of New York, New York, NY 10031, USA}
\affiliation{Kavli Institute for Theoretical Physics, University of California, Santa Barbara, CA 93106, USA}

\author{ Harald~W.~Grie\ss hammer}
\email{hgrie@gwu.edu}
\affiliation{Institute for Nuclear Studies, Department of Physics, George Washington University, Washington DC 20052, USA}
\date{\today}

\begin{abstract}

We find the threshold structure of the two- and three-nucleon systems,
with the deuteron and \tr/\hes~as the only bound nuclei, sufficient
to predict a pair of four-nucleon states: a deeply bound state
which is identified with the \he~ground state, and a shallow,
unstable state at an energy $B^*_\alpha=[0.38\pm0.25]~\text{MeV}$ above the
triton-proton threshold which is
consistent with data on the first excited state of the \he.
The analysis employs the framework of Pionless EFT at leading order with a
generalized regulator prescription
which probes renormalization-group invariance
of the two states with respect to higher-order perturbations
including asymmetrical disturbances of the short-distance
structure of the interaction.
In addition to this invariance of the bound-state spectrum
and the diagonal \tr-$p$ $^1S_0$ phase shifts in the \he~channel
with respect to the short-distance structure of the nuclear interaction, 
our multi-channel calculations with a resonating-group method
demonstrate the increasing sensitivity of nuclei to the neutron-proton $P$-wave
interaction. We show that two-nucleon phase shifts, the triton channel,
and three-nucleon negative-parity channels are less sensitive
with respect to enhanced two-nucleon $P$-wave attraction than the
four-nucleon \tr-$p$ $^1S_0$ phase shifts.

\end{abstract}
\maketitle

\section{Overture}
\la{sec:overt}

The amount of complexity in a system grows with its number of constituents.
It is a challenge for any theory to relate phenomena, which are absent in a
system of $n$ particles but emerge in a $(n+1)$-particle system, to
a set of parameters which characterizes the interaction between only $n$
particles. In nuclear physics, in particular, a useful theory
should, at least, yield shallow two-nucleon states, a stable triton (\tr) and \hes,
and a relatively deeply bound $\alpha$ particle.
By now, one understands to construct such a theory without exchange particles, solely
with neutron ($n$) and proton ($p$) degrees of freedom (pionless effective field
theory~\cite{Kaplan:1998tg,vanKolck:1998bw,Chen:1999tn,Kong2000137} (\eftnopi)).
This theory uses the \tr~binding energy~\cite{Bedaque:1999ve} as
a renormalization condition and
succeeds in the postdiction of a stable ground state
of the four-nucleon system~\cite{Platter2005254}.
It finds the $\alpha$ particle deeply bound and thereby demonstrates
the correlation of a system's complex behavior with properties of its subsystems.
Whether non-bound-state phenomena in complex nuclei are correlated
similarly to two and three-nucleon properties or if these constitute genuine
many-body properties is unknown.

It is the aim of this work to analyze one of these phenomena of the \he~nucleus
in that light: the first excited
$0^+$ state in the ${}^4$He spectrum.
We deem this observable of particular importance because pairs
of a deep ground state and a shallow
excited state with identical quantum numbers reoccur in larger nuclei ($^{12}$C, $^{16}$O) while
in other systems (\eg, $^5$He, $^8$Be)
no stable ground state is sustained below the threshold states.
The missing bound ground state might be a hallmark of the fermion
substructure of the nuclei which becomes relevant only for the latter, while
$^{16}$O, for instance, is amenable to a description in terms of four
interacting bosons (the four $\alpha$'s).
For $^5$He, however, the analogy to five unitary bosons or even an $\alpha$
interacting with a neutron is erroneous -- na\"ively, because of the Pauli principle.
Both treatments, as a five- or two-body system demand
momentum-dependent interactions for the description of shallow states
which are accompanied by an inclusion of radial and/or
angular excitations.
To relate these non-$S$-wave interactions model-independently
to properties of the two- and/or $A$-body interaction is, to our
knowledge, an open problem.
It is in particular not understood if the insensitivity
of low-energy amplitudes such as neutron-deuteron scattering,
or the \he~ground state with respect to $P$-wave components of the
two-nucleon (NN) interaction translates to the excited \he~state, which is
a focus of this work.

A systematic approach is given by \eftnopi, which is an appropriate theory
to analyze possible correlations of
these shell-model characteristics to properties of an underlying interaction.
At its leading order (LO), it
comprises momentum-independent - and thus rotationally invariant
in coordinate space - two and three-nucleon interactions. 
Most regularization schemes in numerical coordinate-space calculations
smear the originally point-like nucleons over some volume and thereby
induce non-zero matrix elements between nucleons
in relative $L>0$ waves. Although this is irrelevant for
two-nucleon observables whose asymptotic states are $S$-wave projections,
asymptotic $A>2$ nucleon states
have non-zero overlap with higher-partial-wave
states on at least one of the relative coordinates.
A renormalization-group (RG) analysis must therefore probe whether 
such a regulator-induced
incorporation of higher-order operators is consistent
with the power-counting of \eftnopi: all amplitudes
which are non-zero because of $P$-wave matrix elements
between nucleons must vanish with the removal of the regulator at LO.

As the admixture of
higher-partial-wave interactions depends on the specific regularization and the tool
which is employed to solve the
few-body problem, we introduce a generalized regulator
in order to assess the sensitivity of
observables with respect to EFT-permitted higher-order
contamination in a LO calculation.
The regulator is specified by 2 parameters:
the customary momentum cutoff $\Lambda$, and
a measure $\eps$ of the strength by which
nucleons are allowed to interact asymmetrically.
This prescription allows for a more comprehensive RG analysis when it is
impractical to vary $\Lambda$
over a broad range, \eg, if numerical tools are limited to certain cutoff values but
can vary the strength of non-central operators with relative ease.

For the problem at hand, we find an excited $0^+$ state of the \he~
insensitive to an RG analysis in $(\Lambda,\eps)$ space.
This conclusion is based on the observed resemblance of the calculated
triton-proton (\tr-$p$) phase shifts with data.
With an iteration of higher-order operators whose coupling strength is smoothly
increased from zero, we affect the attraction in two-nucleon $P$ waves.
In this course, another excited state is introduced while the
ground- and excited state remain invariant.

We begin with a more detailed motivation of the theoretical
problem posed by the excited $0^+$ \he~state.
An ansatz for a solution is given after that, when we introduce a non-central
regulator prescription for the nuclear theory without pions.
A presentation of results obtained with this
technique in the two, three, and four-nucleon systems follows
with an emphasis on their respective sensitivity to unphysical short-distance
distortions of the NN potential with a spin-orbit term.

\section{The four-nucleon $J^\pi=0^+$ channel}
\la{sec:problem}

The Thomas collapse, Efimov's limit cycle, and the Phillips correlation are
representatives of the
fundamental problem of whether microscopic theories are useful for the
prediction of complex features of macroscopic systems.
The correlation between the three and four-nucleon ground states is a rare example
of an $A$-body phenomenon being constrained by observables which
involve less than $A$ particles.
Here, we want to study another complex observable
in the four-nucleon system in order to further the understanding
to what extent non-relativistic,
particle-number-conserving theories can be used to predict many-body
complexity from few-body properties.
Specifically, the binding energy of the \he~ground state $\ba$, which
is large relative to the lowest breakup into a triton with $\bt\approx8.48~$MeV
and a proton, has been related in Ref.~\cite{Platter2005254} numerically
to the $np$ scattering lengths $\as\approx -23.7~$fm, $\at\approx 5.42~$fm and $\bt$.
We ask whether the second $0^+$ state ($\ba^*$)
which is located about $0.4~$MeV in the \tr-$p$
continuum is correlated to the same
observables\footnote{Ref.~\cite{Konig:2016utl}~suggests
such a universal, shallow excitation.}.
Ref.~\cite{Deltuva:2010xd}~finds such resonant states in the four-boson
system interacting via a two-body interaction with infinite scattering length.
The analysis of nuclei entails, in contrast, a large but finite two-body
scattering length $a_2$, while the triton can still be thought of as the
lowest three-body bound state in a finite neutron-neutron-proton spectrum
which is the precursor of the infinite Efimov spectrum to develop
for $a_2\to\infty$.

A shallow bound state emerges
naturally if the interaction between
\tr~and a proton close to threshold is similar to that
of the neutron with a proton close to zero energy:
two particles which are treated as point-like
on the energy and momentum scales involved and
which are subject to very large $S$-wave scattering lengths.
Compared with the neutron, the triton is larger and we expect the
interaction with the proton to be less affected by the repulsive core.
A small increase in the $^1S_0$ attraction would turn the virtual
into a bound state and introduce a new virtual
pole at $E=-\infty$~\cite{newton12} which is far outside of the range
of applicability of \eftnopi.
The original virtual state would become more and more
bound and finally settle to form the \he~ground state while Simultaneously
the newly created virtual state will approach threshold.
There is no {\it a priori} reason why the
virtual state should be close to threshold when the ground-state energy is in an 
EFT-consistent interval around data.
Only if the effective \tr-$p$ interaction which emerges
from the two and three-nucleon operators of
\eftnopi~exhibits this feature, the theory has a chance to converge to the
experimentally found shallow resonance.

\section{Asymmetric Regularization of a non-relativistic theory}
\la{sec:theory}

In the development of contact field theories for nuclei,
cutoff schemes with a regularized delta-function:
\mbox{$\delta_\Lambda^{(3)}(\ve{r})\propto e^{-\Lambda^2/4\ve{r}^2}$,}
provide an intuitive method to renormalize two-body amplitudes in coordinate space.
This spherically-symmetric regulator admits more or less $S$-wave modes in the
calculation of an observable. Useful theories are insensitive to additional
modes which resolve structure below some radial separation. A general RG transformation
would, in addition, assess the effect of small couplings between modes of different
relative angular momentum. Such a generalized analysis is unnecessary for the
low-energy two-nucleon system because a $np$ $S$-wave
state does not couple to higher partial waves at lowest order in \eftnopi.
Amplitudes with more than three particles in the asymptotic states inevitably
involve relative motions with non-zero angular momentum.
A regulator which probes sensitivity with respect to short-distance structure
more comprehensively is apt:
\be\la{eq:gen-reg}
\delta_{\Lambda,\ve{r},\ve{\nabla},\ve{\sigma}_{1,2}}^{(3)}(\ve{r})\propto e^{~\sum_i\frac{\epsilon_i}{\Lambda^{n(i)}}\,\hat{O}_i(\ve{r},\ve{\nabla},\ve{\sigma}_{1,2})}\;\;.
\ee
The general regulator may include an infinite number of higher-order
operators $\hat{O}_i$ with increasing mass dimension $n(i)$ and dimensionless
constants $\eps_i<\infty$.
In this form, it satisfies the condition
\mbox{$\lim_{\Lambda\to\infty}\delta_{\Lambda,\ve{r},\ve{\nabla},\ve{\sigma}_{1,2}}^{(3)}(\ve{r})=\delta(\ve{r})$} and
respects the $SO(3)\otimes SU(2)$ symmetry of
the Hamiltonian through constraints on $\hat{O}_i$.
The customary form is $\hat{O}=\ve{r}^2$ and
$\eps=-\nicefrac{1}{4}$.
In addition,
an explicit dependence on
the relative coordinate between the interacting particles, $\ve{r}$,
the angular momentum associated with this coordinate, $\ve{L}=-i\ve{r}\times\ve{\nabla}$, and
the spin degrees of freedom, $\ve{\sigma}_{1,2}$, is included here.

Explicitly, we choose 
\begin{align}\la{eq:regls}
\delta_{\Lambda,\ve{L},\ve{s}_{1,2}}^{(3)}(\ve{r})\propto\,e^{-\frac{\Lambda^2}{4}\ve{r}^2+\frac{\eps}{2}\,\ve{L}\cdot(\ve{s}_1+\ve{s}_2)}\;\;
\end{align}
and thereby
to analyze sensitivity to spin-orbit distortions because
firstly the two-nucleon LECs do not have to be re-calibrated.
These LO LECs are fitted to $S$-wave observables which are
unaffected by the $\ls$ term
whose effect on higher partial waves also vanishes for $\Lambda\to\infty$
because of the node of the radial wave function at zero distance.
Second, it is the operator of lowest mass dimension which induces transitions between
states of different orbital-angular momentum.
In addition, one has to demand a small size
of the operator matrix element which
characterizes the amplitude of interest.
For nuclear states with good total spin, angular
momentum, and total momentum, for example, the condition
\begin{equation}\la{eq:regcond}
\eps\times\langle\,^{2S'+1}L'_{J}\,\vert\,\ve{L}\cdot\ve{S}\,\vert\,^{2S+1}L_J\,\rangle<1
\;\;,
\end{equation}
must be satisfied in order to use an iterated spin-orbit operator as a regulator.
With $\eps$ subject to this system-specific constraint, standard \eftnopi~is represented by the Hamiltonian:
\begin{align}\la{eq:ham}
  \hat{H}_\text{nucl} = & -\sum_{i}^A \frac{\ve{\nabla}_i^2}{2m} 
                       + \sum_{i<j}^A
                       \left[c^\Lambda_S\,(1-\ve{\sigma}_i\cdot\ve{\sigma}_j)+
                       c^\Lambda_T\,(3+\ve{\sigma}_i\cdot\ve{\sigma}_j+\eps\,\ve{L}_{ij}\cdot\ve{S}_{ij})
                        \right]e^{-\frac{\Lambda^2}{4}\ve{r}_{ij}^2}\nonumber\\
                       &+ \sum_{i<j<k}^A \sum_\text{cyc}
                      d_3^\Lambda\,\Big(\frac{1}{2}-\frac{1}{6}\,\ve{\tau}_i\cdot\ve{\tau}_j\Big)
                         e^{-\frac{\Lambda^2}{4}\left(\ve{r}_{ij}^2+\ve{r}_{ik}^2\right)}\;\;. 
\end{align}
In this form, the spin-orbit interaction can be understood as an {\it irrelevant}
operator, whose contribution vanishes because its LEC $c^\Lambda_T\eps$ is not
renormalized. As for its higher mass dimension, the renormalization of $c^\Lambda_T$
is insufficient, and hence, there must be no contribution to amplitudes from
the $\ls$ term in the limit $\Lambda\to\infty$. 
For finite $\Lambda$ and $\eps\neq0$, the short-distance
behavior of the EFT is distorted asymmetrically\footnote{We use
$\ve{L}_{ij}\cdot\ve{S}_{ij}=\frac{1}{2}(\ve{r}_i-\ve{r}_j)\times(\ve{\nabla}_i-\ve{\nabla}_j)\cdot(\ve{\sigma}_i+\ve{\sigma}_j)$
and calibrate $c^\Lambda_{T,S}$ to the deuteron binding energy $\bd=2.224~$MeV,
the singlet $np$ scattering length $a_s=-23.75~$fm.
In the fit of $d_3$ to $\bt=8.482~$MeV, the effect of
the spin-orbit term is insignificant. Numerical values are given in Table~\ref{tab:input}.}.
We expanded Eq.~\eqref{eq:regls} for practical reasons and use the linear term
of the exponential, only.
This expansion introduces an upper bound for the second RG parameter $\eps$
while in the exponent any $\eps$ is admissible.
With this approximation, Eq.~\eqref{eq:regcond} constrains $\eps$,
the iterated spin-orbit interaction represents
an uncontrolled higher-order contribution.
Following Refs.~\cite{PhysRevC.89.064003,Kirscher:2015zoa}, which show
that the proton-proton Coulomb interaction is a perturbation in light
nuclear bound states, we set $\alpha_\text{EM}=0$.
However, for scattering of charged nuclei with asymptotic center-of-mass momenta
$\lesssim 10~$MeV,
the Coulomb interaction has to be included non-perturbatively~\cite{Kong2000137}.
As the \tr$(p,n)$\hes~reaction is an integral part
of any analysis of the ${}^4$He system, 
we elaborate on its effect when we discuss \tr-$p$ scattering below
in Fig.~\ref{fig:4n-delta-coul}.

With the basic NN $P$-wave phase shifts,
we illustrate the sensitivity of nuclear amplitudes with respect to
either RG parameter in Fig.~\ref{fig:2n-phas}.

\begin{figure}
\includegraphics[width=.49 \textwidth]{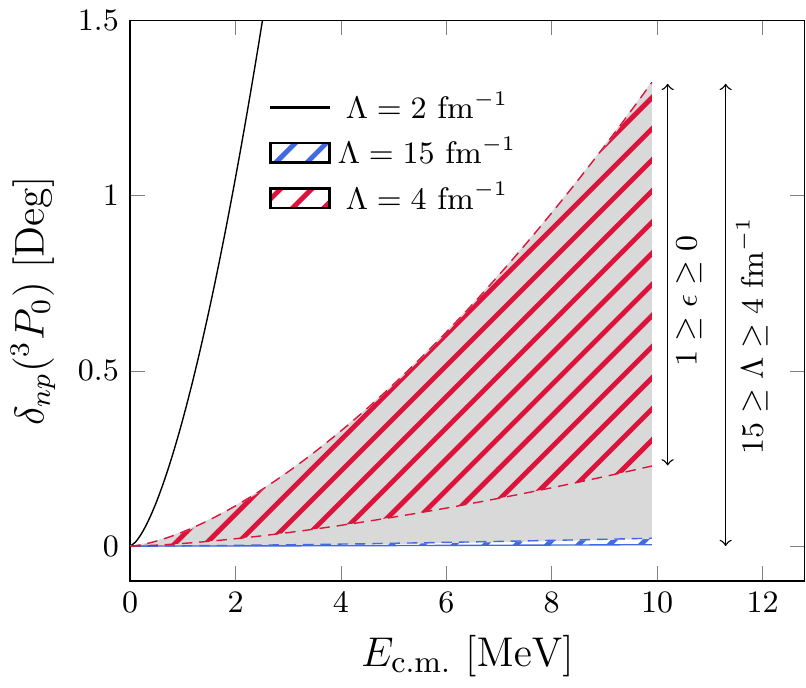}
\caption{\small Energy dependence of the two-nucleon $^3P_0$ phase shifts as a function of the
regulator cutoff $\Lambda\in[4,15]~\text{fm}^{-1}$ (upper and lower edge of the gray area)
and the spin-orbit parameter $\eps\in[0,1]$ for fixed $\Lambda$ ($4~\text{fm}^{-1}$: red hatched,
$15~\text{fm}^{-1}$: blue hatched).
\la{fig:2n-phas}}
\end{figure}

Varying $\Lambda$ between $4~\text{fm}^{-1}$ and $15~\text{fm}^{-1}$ results
in phase shifts within the
gray area. In the zero-range limit $\Lambda\to\infty$,
the interaction does not affect this partial wave.
The phase shifts  converge to zero.
On the other end, the relatively steep rise of the phase for
$\Lambda=2~\text{fm}^{-1}$ (solid black line) does imply a minimal value for the cutoff
in calculations with $P$-wave components in the
asymptotic states which exceeds the na\"ive two-body
breakdown scale. As these higher-partial waves are elements of multi-nucleon states, results
obtained with cutoffs only slightly larger than
$\Lambda\sim0.5~\text{fm}^{-1}$, \ie, the pion-cut scale,
are affected by unphysical poles.

Almost the same phase uncertainty mapped out by this $\Lambda$ variation,
can be parameterized by changing $\eps$.
This is shown by the red (blue) hatched area which results from an $\eps\in[0,1]$
with a fixed $\Lambda=4(15)~\text{fm}^{-1}$.
The effect of the $\eps$ variation diminishes with
increasing $\Lambda$. This is consistent with
a unique zero-range, \ie, $\Lambda\to\infty$, limit:
Observables converge to the same value in this
limit, regardless of the admixture of higher-order
operators through the regulator via Eq.~\eqref{eq:gen-reg}.
A more comprehensive analysis of the sensitivity of
NN $P$-wave observables follows in the discussion
of Fig.~\ref{fig:2n-delta}.

\section{Spin-orbit dependence of $A\leq3$ nuclei}
\la{sec:results}

We first demonstrate the effect of a variation of the spin-orbit strength $\eps$
in the two and three-nucleon sector. To qualify this as analogous to the
conventional probe of sensitivity to high-energy modes via a $\Lambda$ variation,
the effect on two-nucleon $P$-wave systems and \tr~must be parametrically small.
As there is no experimental evidence of shallow poles in the two-nucleon
sector beside the
ones corresponding to the deuteron and its virtual copy in the singlet channel,  
and because the \eftnopi~tenet demands higher-order contributions to be perturbative,
$\eps$ must not create any of these states with typical momenta which are smaller
than the EFT's breakdown scale.

This criterion is validated with the phase shifts as shown in Fig.~\ref{fig:2n-delta}
and complements the result presented in Fig.~\ref{fig:2n-phas}.
In all three $S=1$
NN $P$-wave channels\footnote{$S=0$ matrix elements of the spin-orbit force vanish.},
the considered $\eps$'s induce a spread of the phases ($^3P_0$: blue,
$^3P_1$: gray, $^3P_2$: red) which precludes an emerging pole below 10~MeV.
The attractive character in the $J=0,1$ channels is obscured by the contribution of
the $c^\Lambda_T$ contact term. With this effect removed,
\ie, $c^\Lambda_T=0$, we obtain the less opaque bands shown
in Fig.~\ref{fig:2n-delta}. These bands are now ordered as expected,
$\ve{L}\cdot\ve{S}~|SLJ\rangle=\nicefrac{1}{2}~(j(j+1)-S(S+1)-L(L+1))|SLJ\rangle$.
The interaction
is relatively strong for $J=0$ (faint blue band), and weaker but of similar
significance, albeit of different sign, for $J=1$ and $J=2$.
The band spreads are induced with $\eps\in[0,1]$. At $10$~MeV, the width of
all bands is $<0.1$~Deg and thus neither suggests
a resonance below $\ecm=10~$MeV. Hence, the effect of the spin-orbit distortion in the
two-body sector is as small as required of a regulator.
Emerging two-nucleon bound states with negative
parity indicate that $\eps$ exceeds the regulator range.
This occurs at $\eps\gtrsim1.58$ in the $^3P_0$ $np$ channel as shown
in Fig.~\ref{fig:3n} (blue solid line).
From the appearance of the first bound state for an $\eps\gtrsim1.58$, we can infer that
$\langle\,^3P_0\,\vert\,\ve{L}\cdot\ve{S}\,\vert\,^3P_0\,\rangle=\mathcal{O}(10^{-1})$.

\begin{figure}
\includegraphics[width=.49 \textwidth]{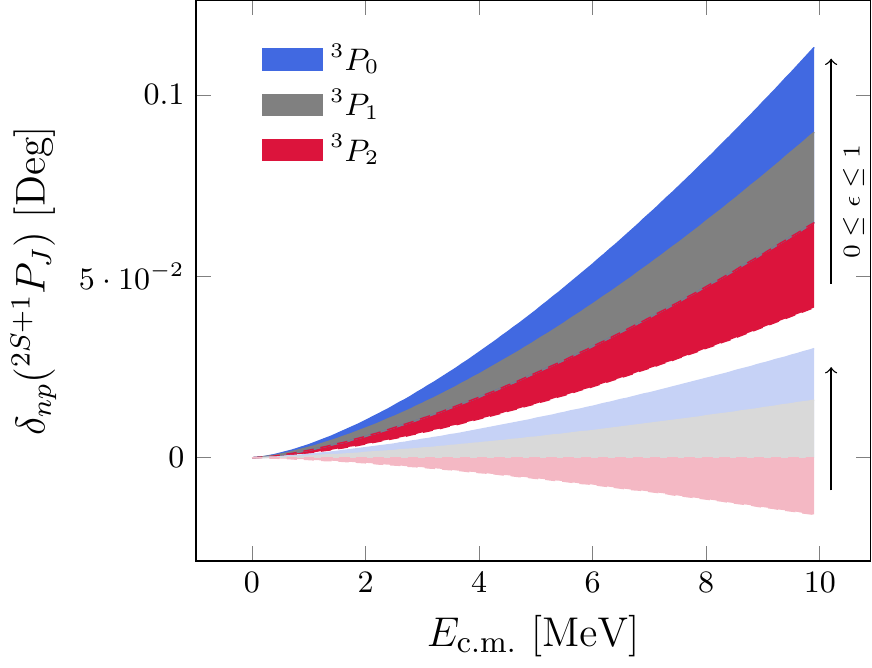}
\caption{\small Energy dependence of the two-nucleon $P$-wave phase shifts with
regulator \lam{6}. The bands result from a variation of the
spin-orbit strength in the interval $[0,1]$. Lightly shaded bands were obtained with $c_{T,S}^\Lambda=0$.
\la{fig:2n-delta}}
\end{figure}

While the two-nucleon bound state (deuteron) contains only orbital $S$-wave components
at LO in \eftnopi, nucleons in relative $L>0$ states are part of the three-nucleon $S=\frac{1}{2}$
bound state (triton). 
In a properly renormalized EFT, these non-$S$-wave states which contribute, here,
to observables via the $\ls$ regulator and the finite cutoff, add to the overall
theoretical EFT uncertainty. They
must vanish in the $\Lambda\to\infty$ limit.
The \eftnopi~expansion of the two-nucleon scattering amplitude, in particular, is
ordered such that relative $P$-waves become relevant at $\mathcal{O}(a_{s}^{-3}/m^3)$,
and thus we assume that ensuing non-zero effects of the
spin-orbit distortion on a three-body bound state are also suppressed relative to the
LO part of the interaction.
To validate this claim, the dependence of eigenvalues of the Hamiltonian
Eq.~\eqref{eq:ham} on $\eps$ with $\Lambda=6~\text{fm}^{-1}$ and LECs
as given in Table~\ref{tab:input}, was calculated and results
are shown in Fig.~\ref{fig:3n}.

For $\eps\lesssim1.5$, the only eigenvalue (solid red line) below the
deuteron-neutron ($dn$)
threshold (black solid line) is $\bt$, \ie, the spin-orbit distortion has
no significant effect on \tr~as calibrated at LO in \eftnopi.
An excited state emerges out of the $dn$ continuum and settles at $\bt$ for
\mbox{$1.5\lesssim\eps\lesssim1.52$}. The gap between $\bt$ and the ground state
becomes very large relative to \mbox{$\bt-\bd$}, and \tr~could be identified with
an excited state since the effects of the deep ground state are small.

three-nucleon states with negative parity have non-zero overlap with four-nucleon $0^+$ states.
Although, their contribution to the \he~ground state is small, it is imperative to analyze the
two negative-parity
$d-n$ channels: $\frac{1}{2}^-$ and $\frac{3}{2}^-$, for their response to a $\eps$ variation.
No bound states have been found in these channels experimentally, and
\eftnopi~with $\eps=0$ does not sustain them. 
They are expected to appear at a critical $\eps$ strength, which
is given in Fig.~\ref{fig:e_critical}.
We compile in Fig.~\ref{fig:e_critical} the $\Lambda$ dependence of
critical $\eps$ values for
all $A=2,3$ observables we analyzed.
The values are defined such that for $\eps\approx\eps_c$ an
unphysical bound state enters the
respective spectrum at threshold.
For all negative-parity states, this new state is the
ground state. In the triton
channel, $\eps_c$ indicates the strength at which a second state
becomes bound at the $dn$ threshold.
The ordering of the two-body $\eps_c$'s corresponds to the
$J$-dependent strength of the $\ls$ interaction in the
respective channel, \eg, the smallest
$\eps_c$ is found in the channel where the
interaction is strongest, \ie, $^3P_0$ (blue line in
Fig.~\ref{fig:e_critical}).
All critical strengths, for the three two-nucleon channels, the triton (solid green)
and the two negative-parity three-nucleon channels
($\nicefrac{1}{2}^-(\nicefrac{3}{2}^-)$): green(red) band),
converge with the cutoff regulator $\Lambda$.

For all channels, we find $\eps_c\gtrsim1$.
Critical values of the two-nucleon system are larger than for three nucleons.
This follows from the larger $\ls$-operator matrix elements in Eq.~\ref{eq:regcond}
when in $A>2$ systems more than just one pair of nucleons interacts.
The ratio $\eps_c(2n,{}^3P_0)/\eps_c(3n,{}^2P_1)\sim1.5$ follows from applying the
pair-counting formulas of Ref.~\cite{Wiringa:2006ih}~if we assume that the spatial
symmetry of the three-nucleon state is given by the Young diagram $[21]$. 
A generalization of this method could be used to predict critical strengths in larger
systems from those of their subsystems by counting the number of pertinent pairs.
Counting the spin 1, isospin 0 pairs in
\he~suggests $\eps_c(4n,{}^1S_0)\approx2\eps_c(3n,{}^2S_1)$
which is very close to the numerical value as discussed in Sec.~\ref{sec:4n}. 

\begin{figure}
\includegraphics[width=.49 \textwidth]{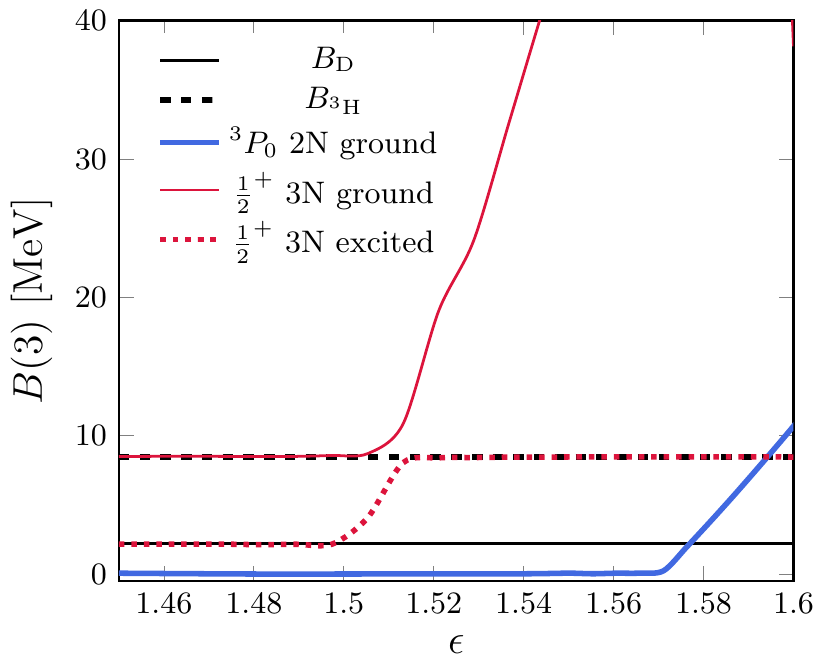}
%
\caption{\small Energy eigenvalues in the three-nucleon $J^\pi=\frac{1}{2}^+$ (solid red: ground state; dotted: first excited state) and two-nucleon ${}^3P_0$ (solid blue)
channels as a function of the spin-orbit component of the regulator for
$\Lambda=6~\text{fm}^{-1}$.
Solid and dashed horizontal lines are drawn at $\bd$ and $\bt$.
\la{fig:3n}}
\end{figure}

\begin{figure}
\includegraphics[width=.49 \textwidth]{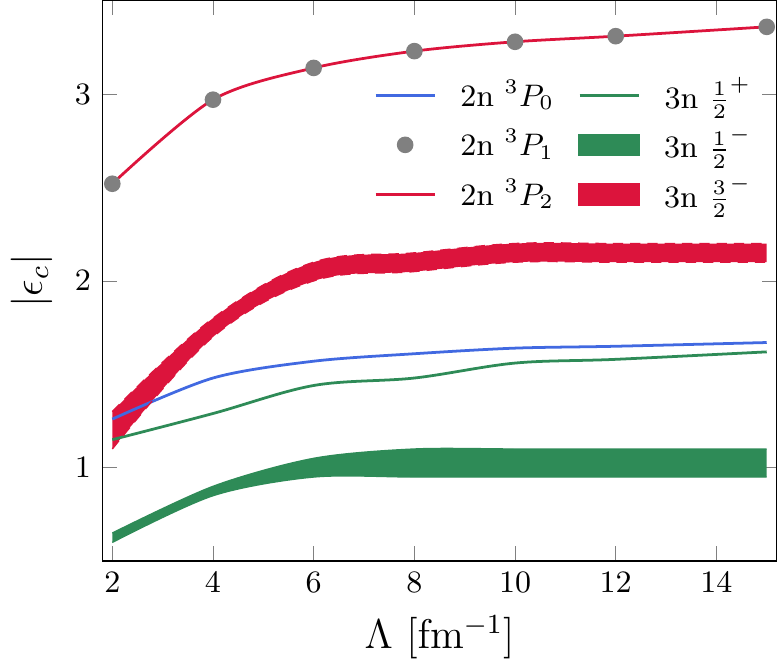}
\caption{\small 
$\Lambda$-cutoff dependence of critical strengths of the non-central regulator
component $\eps_c$, above which additional bound states appear.
The dependency is shown for the two-nucleon $P$-wave channels $^{2S+1}P_J$,
the three-nucleon $\nicefrac{1}{2}^+$ triton channel (solid green), and the three-nucleon negative-parity
channels $\nicefrac{1}{2}^-(\nicefrac{3}{2}^-)$ (green(red) areas whose width represents numerical uncertainty).
\la{fig:e_critical}}
\end{figure}

Finally, it is noteworthy that the $\nicefrac{1}{2}^-$ three-nucleon channel
is more sensitive to a spin-orbit distortion compared with
the triton $\nicefrac{1}{2}^+$ ground-state channel. This has to be considered
in scenarios where the spin-orbit force is physical, \eg, of a nucleus is
subjected to external fields. Consider, for instance, a $nnp$ halo in the
effective potential of a core nucleus. A state for which this might be
relevant is the first excited state in lithium-9 with $J^\pi=\nicefrac{1}{2}^-$.
These quantum numbers are consistent with a helium-6, $\nicefrac{1}{2}^-(nnp)$
cluster structure. Because of the particle instability of the negative parity
three-nucleon states in vacuum, such a configuration is, to our knowledge, not
of major focus when these system is studied. Here, we conclude qualitatively that
the effective interaction between the helium-6 core and the three nucleons favors
their $\nicefrac{1}{2}^-$ configuration.

\section{4 nucleons}
\la{sec:4n}

In the two- and three-nucleon channels considered above, there is no
data which indicates a shallow resonant state.
The $\alpha$ particle is thus the smallest nucleus which sustains such a
state close to its lowest break-up threshold.
The RG invariant existence of such a state at LO in \eftnopi~is subject of the
following analyses of the $\eps$ sensitivity of the four-nucleon
{\it bound} (Fig.~\ref{fig:4n}) and {\it scattering} (Fig.~\ref{fig:4n-delta}) systems.

\parg{Bound-state spectrum}
\la{sec:4n-bs}
The effect of the spin-orbit interaction on the $P$-wave components in
the \he~channel is shown in Fig.~\ref{fig:4n}. The dependence of the eigenvalues
of the $A=4$ Hamiltonian on $\eps$ is similar to the triton channel (see Fig.~\ref{fig:3n}).
Specifically, we find that below
a $\Lambda$-dependent critical value, the calculated spectrum contains the ground
state (solid lines\footnote{From $\Lambda\in[4,12]~\text{fm}^{-1}$,
a polynomial extrapolation yields
$\lim_{\Lambda\to\infty}\ba\sim28.9~$MeV which is approached from below.}
in Fig.~\ref{fig:4n})
and an accumulation of eigenstates whose energy cannot be discriminated numerically from $\bt$.
At the critical spin-orbit strength, an excited $0^+$ state (dashed lines) below
the \tr-$p$ threshold emerges. The observed values,
$\eps_c(\text{\lam{4}})\sim0.58$, $\eps_c(\text{\lam{6}})\sim0.67$, and
$\eps_c(\text{\lam{8}})\sim0.72$ suggest almost the value predicted above with
pair counting: $\lim_{\Lambda\to\infty}\eps_c\sim0.9$
(see the intersect of the respective blue, red, and green dashed lines in Fig.~\ref{fig:4n}
with the black dashed line). The limit is near the point where a three-nucleon $\nicefrac{1}{2}^-$
state becomes bound. The excited state's structure is accordingly a
proton orbiting in a $P$-wave around a negative parity triton. In comparison, the dominant
structure of the ground state is a proton in a $S$-wave relative to a \tr~core.
The $\eps_c$ in the four-body system is smaller compared with the critical values in the
2, and three-body systems for $\Lambda<\infty$.

These results are consistent with the interpretation of the iterated spin-orbit term
as part of \eftnopi~for $\eps<\eps_c$.
The increasing $\ls$ sensitivity of nuclear systems with their particle number
stems from the matrix element of
the spin-orbit operator (see Eq.~\eqref{eq:regcond}). The contribution
of a two-nucleon pair in an eigenstate of the spin-orbit operator to the energy of an
$A$-body state is expected to scale with $A$. Therefore, any attractive spin-orbit interaction
will eventually bind an $A$-body nucleus, regardless of how small it is in the two-nucleon
system.
In practice, the different $\eps_c$ pertinent to
the two, three, and four-body observables allow for a parameterization of an interaction
which describes consistently the two, three, and four-nucleon system, like \eftnopi, with a handle on the
character of a shallow $0^+$ state in the $\alpha$ channel.

\begin{figure}
\includegraphics[width=.49 \textwidth]{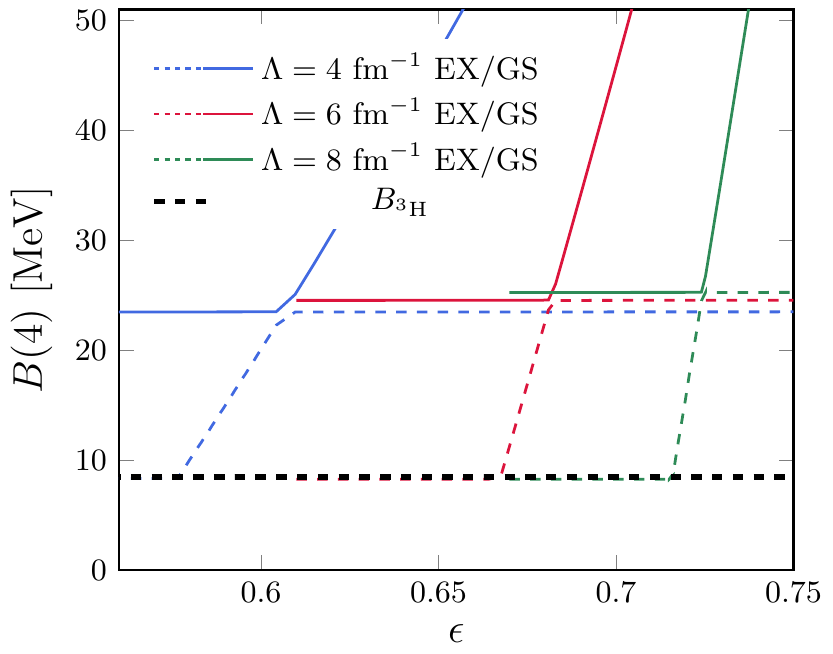}
\caption{\small Energy eigenvalues (Ground State: solid; EXcited state: dashed) in
the four-nucleon $J^\pi=0^+$
channel as a function of the spin-orbit component of the regulator for \lam{4} (blue), \lam{6} (red), and
\lam{8} (green).
A dashed horizontal line indicates the \tr$-$nucleon threshold.
\la{fig:4n}}
\end{figure}

\parg{Elastic scattering}
\la{sec:4n-sc}
The four-nucleon bound states emerge for $\eps$ strengths which are
compatible with an interpretation of the spin-orbit distortion as a
regulator effect in the two- and three-nucleon systems. Therefore, we
cannot rule out that these states evolve from former physical resonances,
and that the $\eps$ dependence indicates the need for a modified EFT
power counting.
In this section, we refute this possibility. We demonstrate that besides
the excited states which emerge with increasing $\eps$ another, $\eps$ invariant
$0^+$ resonance exists which we identify with the physical state.
The character of the latter cannot be inferred from the spectrum as
obtained in
Fig.~\ref{fig:4n}\footnote{See Refs.~\cite{Stetcu:2006ey,Konig:2016utl}
for other investigations which report on an excited shallow state.}.
The LO \eftnopi~analysis of the $^4$He scattering
system, below, provides comprehensive evidence for both, the existence of
a resonant state, and its RG invariance in contrast to the excitations
which exhibit such an $\eps$/RG dependence.

The results are shown in Fig.~\ref{fig:4n-delta}.
We focus on energies around the \tr-$p$ threshold where experiment locates
the excited $\alpha$ state. In this interval, we consider two coupled
channels, \tr-$p$ (solid phase shifts) and \hes-$n$ (dashed phase shifts).
We remove the degeneracy of these
channels which follows from the absence of strong- and electroweak iso-spin
dependent components at LO \eftnopi~(\cf~Eq.~\eqref{eq:ham}) by shrinking
the variational basis for ${}^3\text{He}$. This enforces different boundary
conditions, namely for a neutron sufficiently far from the ${}^3\text{He}$,
the latter is stable with an energy approximately $0.6$~MeV less than $\bt$.
We do thereby not change the variational basis (see also Appendix~\ref{sec.app1})
which is used to approximate the wave function in a region where the particles
interact strongly. Resonant behavior is thus modeled is the same way as it is
for degenerate asymptotic states,
but effects on the phase shifts from the channel coupling are disentangled
from that of the resonant states.

We first discuss diagonal phase shifts (the off-diagonal phases/mixing angles are
analyzed in the Appendix) as shown in Fig.~\ref{fig:4n-delta}~for
$\Lambda\in[4,12]~\text{fm}^{-1}$ (color-coded)
with $\eps$-regulator parameters outside an epsilon surrounding of the
critical values, $|\eps-\eps_c|\gtrsim10^{-2}$.
The phases rise from zero energy (relative to $\bt$) up to a discontinuity
before a cusp
marks the opening of the \hes-$n$ channel (barely visible kink at, \eg,
$0.6~$MeV for \lam{8}~at the onset of the dashed line).
The shape of the discontinuity is characteristic~\cite{newton17}~of
a threshold behavior in a
multi-channel problem. We conclude from the absence of other particle-stable
nuclei with $A\leq4$ besides the ones which are included in our calculation
that the spikes correspond to resonant states.

The position of the resonance/discontinuity moves away from
threshold with increasing $\Lambda$. This motion correlates with the
deliberately chosen gap between the two open channels which widens
also with increasing $\Lambda$. The relative
position of the spike between zero energy and the threshold of the \hes-$n$
channel is visibly the same for the three $\Lambda$ values.
Hence, we postdict the physical resonance with $J^\pi=0^+$ between
the \tr-$p$ and \hes-$n$ threshold independently of $\Lambda$.
We quantify the prediction by first extrapolating from the three
calculated cutoff values the limit
\be\la{eq:eres-by-egap}
\lim_{\Lambda\to\infty}\frac{B^*_\alpha}{\bt-\bh}=0.5\;\;\;.
\ee
Now, we choose to extract the EFT prediction for $B^*_\alpha$ where the
experimental value for the energy difference between the two thresholds is
reproduced exactly, namely: $\bt-\bh\approx0.76~$MeV, and we obtain
\be\la{eq:respos-pred}
B^*_\alpha\big(\eftnopi\big)=[\,0.38\pm0.15_\text{\tiny EFT}\pm0.1_\text{\tiny\rgm}\,]~\text{MeV}\;\;.
\ee
The EFT uncertainty is inferred from the difference of the ratio in
Eq.~\eqref{eq:eres-by-egap} between \lam{4}~and \lam{12}. It is
consistent with an a-priori estimate $\nicefrac{1}{3}$ of the LO result,
using the canonical estimate for the expansion parameter of \eftnopi.
The \rgm~uncertainty represents a conservative estimate of the numerical
method. With this uncertainty, the \eftnopi~prediction is consistent
with the experimental value:
$B^*_\alpha\big(\text{exp.}\big)= 0.395(20)~$MeV~\cite{TILLEY19921}.

In addition to the invariance with respect to spatially symmetric
transformations as parameterized by $\Lambda$, we investigate the
sensitivity of the resonance with respect to the spin-orbit distortion. 
The phase shifts as shown in Fig.~\ref{fig:4n-delta} do not change
significantly for any $|\eps-\eps_c|\gtrsim10^{-2}$.
Only if the spin-orbit distortion is tuned to a critical value,
spikes, similar to the displayed ones, indicate the presence of the
negative-parity sub-threshold state. We refrain to display these
additional discontinuities in the figure because the shown spikes and
the associated states which
cause the steep rise of the phases are unaffected,
\ie, phase shifts for
$|\eps-\eps_c|\gtrsim10^{-2}$ are indistinguishable from $\eps=0$.
It is the physical $0^+$ excited state which causes the rise.
From this stability of the phase shifts and the $\Lambda$ invariance,
we conclude that this state is RG invariant. In other words, an excited
four-nucleon state in the \he~channel with $J^\pi=0^+$ is well described
by \eftnopi, and as such correlated with three low-energy data points, \eg,
the deuteron binding energy, the $np$ singlet scattering length, and the
$\bt$.

Below the \hes-$n$ threshold, we identify a second discontinuity
for each cutoff (spikes of dashed/solid lines). These discontinuities are
found at the same energy in the \tr-$p$ phase shifts and indicate
another threshold. Although, a $0^-$ resonance, approximately $0.42$~MeV below the
\hes-$n$ threshold is well established~\cite{TILLEY19921},
this is not the state responsible for the calculated spikes here. We rule
out this possibility by choosing $\eps\approx0$, thereby turning off
the coupling between positive and negative parity states at \lam{12}.
Hence, any effect of a negative-parity state on the considered scattering
problem which defines asymptotic states of positive parity is absent by
construction. As the spikes do not disappear at $\eps=0$,
we interpret the second set of discontinuities as the iso-spin mirror
of the first $0^+$. A possible explanation for its presence is the absence of
the Coulomb interaction in the proton-proton system in our analysis.
If this state has largest overlap with
a \tr-$p$ configuration, the repulsion would further destabilize it and thereby
diminish its effect on the phases in the energy range considered here.
We test this conjecture by including the Coulomb interaction non-perturbatively.
All other numerical and physical parameters are retained. With its full
strength, the effects of the Coulomb interaction dominate the phase shift
behavior. To study the essence of the effect without making the identification
of resonant behavior impractical, we increase the Coulomb interaction gradually
from zero strength by modifying the fine structure constant
to $\kappa\times\alpha_\text{EM}$ and take $\kappa\in\lbrace0.02, 0.4, 1\rbrace$.

The result is shown in Fig.~\ref{fig:4n-delta-coul}
for \lam{4}. Compared with the Coulomb-less
results (blue lines in Fig.~\ref{fig:4n-delta}) the results are qualitatively
similar below $\approx 1~$MeV. Namely, for both attenuation factors ($0.02$ (black)
and 0.4 (gray)), we observe a steep rise of the \tr-$p$ phase shifts at
threshold, which we defined as zero energy. A kink signals the \hes-$n$
threshold. For the weaker Coulomb repulsion, this kink is hardly visible
but apparent for the stronger repulsion where we identify it as a feature
independent of the initial rise, \ie, the $0^+$ resonance.
In contrast to the Coulomb-less phases, the resonant rise in the \hes-$n$
phases cannot be found at a comparable energy to the one in the \tr-$p$
channel. The small hump which can be observed for both Coulomb strengths
in the \hes-$n$ phases can be interpreted as the remnant of a threshold to a
stable or resonant state. Adopting the latter, it would correspond to the
explanation given above, of a resonant state with a dominant triton-proton
component, which is close to threshold if its constituents do not repel each
other, but which becomes increasingly unstable with this repulsion.

\begin{figure}
\begin{minipage}[t]{0.45\linewidth}
\includegraphics[width=.95 \textwidth]{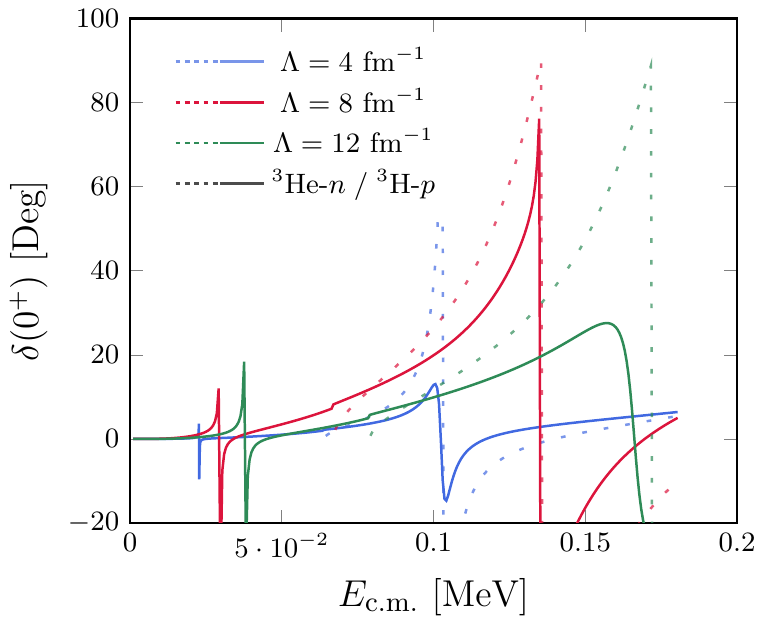}
%
\caption{\small Energy dependence of the diagonal \tr-$p$ (solid) and \hes-$n$
(dashed) scattering phase shifts in the $J^\pi=0^+$ \he~channel
 for \lam{4} (blue, $\eps=0.1$),
\lam{8} (red, $\eps=0.7148$), and \lam{12} (green, $\eps=0.01$).
\la{fig:4n-delta}}
\end{minipage}\qquad\qquad
\begin{minipage}[t]{0.45\linewidth}
\includegraphics[width=.95 \textwidth]{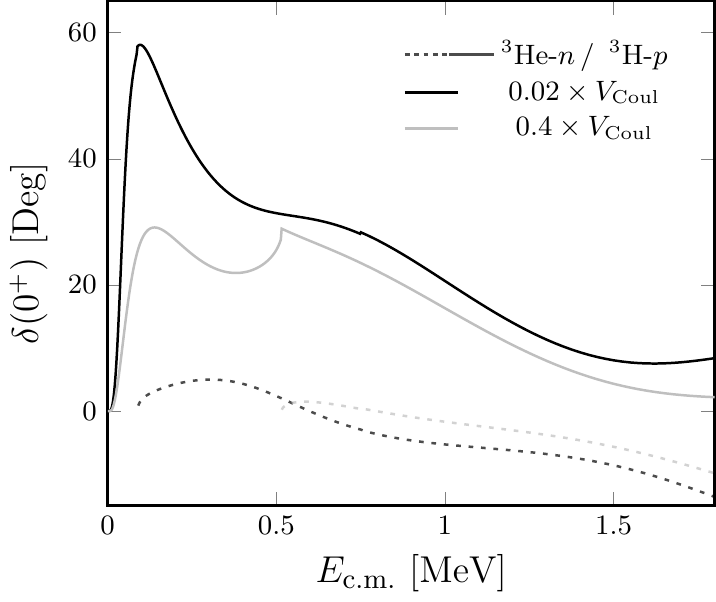}
%
%
\caption{\small \tr-$p$ (solid) and \hes-$n$ (dashed) scattering phase shifts as
in Fig.~\ref{fig:4n-delta} for \lam{4} {\bf with} a scaled (attenuation factors
0.02 (black) and 0.4 (gray)), non-zero Coulomb interaction.
\la{fig:4n-delta-coul}}

\end{minipage}
\end{figure}%

\section{Epilogue}
\la{sec.epi}

We find that the first excited, unstable state of the \he~can be predicted
solely from characteristics of the two- and three-nucleon
subsystems: the deuteron, the virtual singlet neutron-proton state,
and the triton.
The energy of the state is approximately 0.38~MeV above the triton-proton threshold,
and its behavior as a shallow resonance is found independent of a set of
regulator types for two- and three-nucleon contact interactions which
comprise the pionless formulation of a microscopic nuclear theory.
We thus show that this theory does not only correlate the \he~ground state
to properties of its subsystems, but that the pair of a relatively deeply bound
\he~and an unbound excited state is a consequence of an almost unitary
two-body system and a fine-tuned three-body bound state.

The implied renormalization-group invariance was
probed with a generalization of the regulator function.
This enables the assessment of the sensitivity of observables
with respect to asymmetric distortions of the
short-distance structure of the interaction.
This generalized regulator explains the {\it independence}
of low-energy $A\leq4$ observables from
spin-orbit interactions of a certain strength.
The calculated {\it dependence} of $A\leq4$ observables on
relatively strong, two-body spin-orbit interactions does suggest
a correlation between a negative-parity three-nucleon state and four-nucleon systems.

Scattering phase shifts in the \he~channel around the 
\tr-$p$ and \hes-$n$ thresholds were calculated
to identify near-threshold states in
the \he. These calculations represent the first application of \eftnopi~to
this scattering system and thereby extend the usefulness of the theory to
more complex phenomena besides ground-state properties.
Amongst the two near-threshold states, we find the physical, positive parity state
invariant with respect to the generalized regulator.

\section*{Appendix I: The \rgm~calculation and interaction parameters}
\la{sec.app1}

All observables in this work represent
solutions of the stationary Schr\"odinger equation using
the Hamiltonian in Eq.~\eqref{eq:ham}. We obtain them with
a method based on the concept of resonating groups of particles
(original idea:~\cite{Wheeler:1937zza,Wheeler:1937zz};
specific implementation (\rgm):~\cite{Hofmann:1986,Kirscher:2015ana}).
We specify only the most complicated calculation of this work, namely,
four-nucleon scattering.

In the resonating-group ansatz for the wave function
\be\la{eq:rgwf}
\Psi=\mathcal{A}\left\lbrace\sum_i\phi_I^{(i)}\phi_{II}^{(i)}F^{(i)}(\ve{R}_i)\right\rbrace
\ee
we consider components corresponding to a fragmentation of the four-body system into
\tr-$p$, \hes-$n$, and deuteron-deuteron ($d$-$d$).
Each fragment wave function $\phi$
has appropriate $j_{I/II}^\pi$ quantum numbers coupled from total spin $S$
and angular momentum $L$. The two fragment spins are coupled to a channel spin with
values which allow for its coupling with the angular momentum of the relative motion
between the fragments, $F(\ve{R})$, to $J^\pi=0^+$. 
The $\phi$'s follow the (LS)J coupling
scheme. The deuteron resides in a pure relative $S$-wave ($L=0$), while we 
consider components in the \tr/\hes~wave function
 which carry an angular momentum of $L=1$
on both Jacobi coordinates.
 With these constraints, we consider all intermediate couplings
and expand the radial dependencies in a Gaussian basis.
 The width sets for the intermediate
$(s_1s_2)^{s_{12}=1}$ coupling within \tr, \eg, is obtained from a 20-dimensional
$S$-wave deuteron.

The scattering problem is solved by ``freezing'' the wave functions
 of the fragments for
those channels $i$ in Eq.~\eqref{eq:rgwf} which
 are needed to define the asymptotic behavior.
For a nucleon impinging with less than 10~MeV on a \tr~or \hes, we thus include
as open $S$-wave channels: \tr-$p$, \hes-$n$, and $d$-$d$.
 In these cases, the free relative motion
$F$ is expanded in a set of 20 Gaussians in the
 region where the effective fragment-fragment
interaction is non-zero.
 To account for any deviation from this free motion within the
interaction region, we add products of
 single components of the fragments with square-integrable
relative function $F$. We build a complete basis
 in this way until the lowest eigenvalues in
the resulting spectrum are stable at the 100~keV level.

The effect on the \he~of channels which couple negative parity \tr/\hes~via a
relative $P$-wave to $0^+$ was probed and found sufficiently accounted for by
the $(l_1l_2)^{L=0,1,2}$ components of \tr. As these negative-parity
three-nucleon states where also not part of the study by themselves,
we abstained from including states which resemble this coupling scheme in the four-nucleon calculations.

\begin{table}[h]
\renewcommand{\arraystretch}{1.2}
\begin{tabular}{l@{\extracolsep{.4cm}}c@{\extracolsep{.4cm}}c@{\extracolsep{.7cm}}c@{\extracolsep{.7cm}}}
$\Lambda~[\text{fm}^{-1}]$ & $c^\Lambda_T$ & $c^\Lambda_S$ & $d^\Lambda_3$ \\
\hline
$2.00$ & $ -142.364$ & $ -106.279$ & $    68.4883$ \\
$4.00$ & $ -505.164$ & $ -434.958$ & $   677.799 $ \\
$6.00$ & $-1090.58$ &  $ -986.252$ & $  2652.65  $ \\
$8.00$ & $-1898.62$ &  $-1760.16$  & $  7816.23  $ \\
$10.0$ & $-2929.28$ &  $-2756.69$  & $ 20483.2   $ \\
$12.0$ & $-4182.37$ &  $-3975.62$  & $ 50939.9   $ \\
$15.0$ & $-6479.58$ &  $-6221.50$  & $195570.    $ \\
\hline
\hline
\end{tabular}

\caption{\la{tab:input}Numerical values of low-energy constants in MeV for \eftnopi~calibrated to
$\as\approx -23.8~$fm, $\bd=2.22~$MeV, and $\bt=8.48~$MeV.}
\end{table}

\section*{Appendix II: Multi-channel effective-range expansion}
\la{sec.app2}

The discussion of Figs.~\ref{fig:4n-delta} and~\ref{fig:4n-delta-coul}~
above concentrates on the behavior of
the diagonal part of the S-matrix. Below, we analyze the off-diagonal
elements, \ie, how strongly do the two channels couple as a function
of energy?
It is instructive to begin the discussion with an analogy.
The deuteron is a bound state in which the two constituents -- neutron and
proton -- move ``most of the time'' with zero angular momentum relative to each
other. This is characteristic for the nuclear interaction which favors
spatial rotational symmetry by a relatively weak coupling between the
two-nucleon ${}^3S_1$ and ${}^3D_1$ states.
For the numerical values of the standard parameters for the corresponding
two-channel scattering matrix, this weak coupling of angular-momentum
channels translates to a relatively small
mixing angle $\epsilon$ and relatively small eigenphase shifts, describing
asymptotic two-nucleon states in a relative $D$-wave, compared with those
eigenphases which parametrize zero-angular-momentum scattering. For the
generic two-channel S-matrix, we adopt the following standard:

\begin{align}\la{eq:smat}
S=\begin{pmatrix} \eta_{11}e^{2i\delta_{11}} & \eta_{12}e^{2i\delta_{12}} \\ \eta_{21}e^{2i\delta_{21}} & \eta_{22}e^{2i\delta_{22}}\end{pmatrix}=
\begin{pmatrix} \cos\epsilon & -\sin\epsilon \\ \sin\epsilon & \cos\epsilon\end{pmatrix}
\begin{pmatrix} e^{2i\delta_\alpha} & 0 \\ 0 & e^{2i\delta_\beta}\end{pmatrix}
\begin{pmatrix} \cos\epsilon & \sin\epsilon \\ -\sin\epsilon & \cos\epsilon\end{pmatrix}\;\;.
\end{align}

In our analysis, we do not consider channels which differ by the angular momentum
of the relative motion but in the nuclear composition of the fragments.
As we consider the four-nucleon scattering process
in the vicinity of the triton-proton threshold, only two
two-fragment channels -- \tr-$p$ and \hes-$n$ -- are energetically accessible.
It is helpful to think of the \tr-$p$ arrangement as the analog of the $S$-wave'',
and \hes-$n$ as the $D$-wave analog of the deuteron.
The existence of the excited, resonant state was deduced from
the diagonal phases $\delta_{11/22}$, as shown
in Fig.~\ref{fig:4n-delta} and~\ref{fig:4n-delta-coul}, only, without
need to investigate the strength parameters $\eta$. Row and column indices
specify the asymptotic states, $\delta_{12}$ parametrizes, \eg, the probability to
detect a free \hes-$n$~state with relative energy $E_{ch}$ if a proton hits a triton
with a corresponding $E'_{ch}$.

When we do include the coupling strengths
$\eta$ -- in Fig.~\ref{fig:4n-delta-eigen} we use the more common parameterization
with eigenphases $\delta_{\alpha/\beta}$ and mixing angle $\epsilon$ --
we notice two peculiarities: first, in contrast to the mixing
of angular momentum channels in the deuteron ground state, the eigenstates of the
four-nucleon process are superpositions of the \tr-$p$ and \hes-$n$ configurations
with similar weight -- if a neutron impinges on a proton in a relative $S$-wave, it
is very likely to emerge in an $S$-wave while the probability of it being deflected
into a $D$-wave is small; the probability to detect \hes~and a free
neutron after the collision of a neutron with \tr, in contrast,
is almost as high as an elastic reaction.
In terms of S-matrix parameters: the mixing angle (red) rises quickly to a value
close to $\nicefrac{\pi}{4}$ in Fig.~\ref{fig:4n-delta-eigen}.
Second, at an energy between 0.15~MeV and 0.2~MeV, the two channels decouple
and scatter elastically. The scattered waves are not
phase shifted ($\delta_{\alpha/\beta}\approx 0$) and the nucleons
are not rearranged ($\epsilon\approx 0$), \ie, at this energy, the nuclei scatter
elastically like classical particles.
In the vicinity of this peculiar point, the phase shifts behave with
energy reminiscent of the dependence of energy eigenvalues of a two-level
system around an avoided level crossing. Here, we find the effect 
of the scattering process on one eigenstate insignificant 
(gray phase for $E\lesssim0.17~$MeV and
black phase for $E\gtrsim0.17~$MeV) compared to that on the other eigenstate.
It is beyond the scope of this study to analyze the sensitivity of that phenomenon
with respect to cutoff variations and the proton-proton Coulomb interaction.
Finally, note the cusp in the mixing angle around the energy where we
identified the resonant state from the diagonal phases (red, dashed line
at $\approx0.9~$MeV).

\begin{figure}
\includegraphics[width=.49 \textwidth]{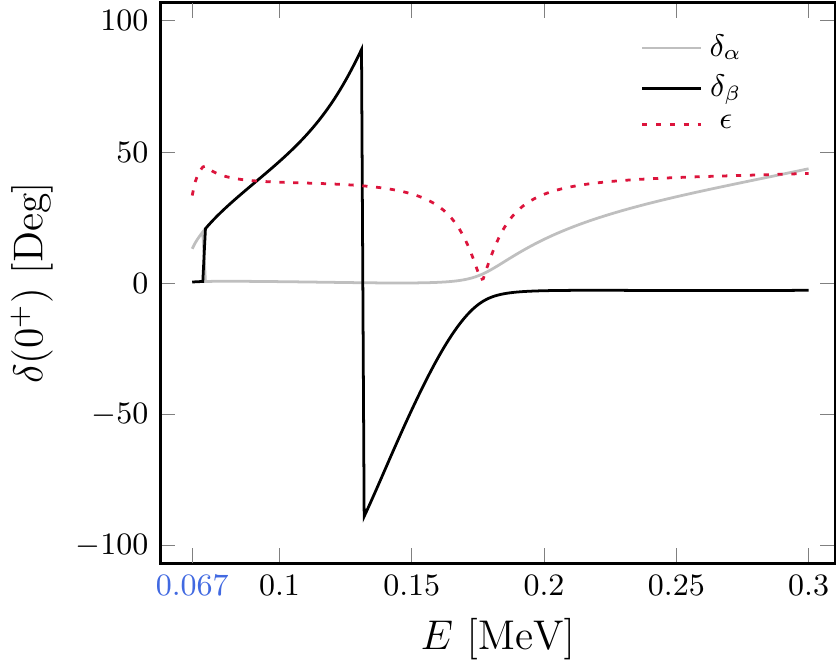}
%
%
\caption{\small Eigenphases (black/gray) and mixing angle (red, dashed)
for the coupled \tr-$p$, \hes-$n$
two-channel scattering system ($J^\pi=0^+$). The energy is defined relative to the
\tr-$p$ threshold (blue abscissa label at 0.067~MeV), \lam{8},
and $\alpha_\text{EM}=0$.
\la{fig:4n-delta-eigen}}
\end{figure}

\section*{Acknowledgements}

Discussions with D.~Lee and B.~Tiburzi, and the hospitality of the KITP and
the University of Trento are gratefully acknowledged.
Comments on the manuscript by W.~Leidemann and G.~Orlandini were of special value.
HWG was supported by the US Department of Energy under contract DE-SC0015393,
and by the Dean's Research Chair programme of the Columbian College of Arts and
Sciences of The George Washington University, and
JK by the National Science Foundation under
Grant No. NSF PHY-1125915, PHY15-15738, and PHY-1748958.

\bibliographystyle{hieeetr}
\bibliography{refs}

\begin{thebibliography}{10}

\bibitem{Kaplan:1998tg}
D.~B. Kaplan, M.~J. Savage, and M.~B. Wise, ``{A New expansion for
  nucleon-nucleon interactions},'' {\em Phys. Lett.}, vol.~B424, pp.~390--396,
  1998, nucl-th/9801034.

\bibitem{vanKolck:1998bw}
U.~van Kolck, ``{Effective field theory of short range forces},'' {\em Nucl.
  Phys.}, vol.~A645, pp.~273--302, 1999, nucl-th/9808007.

\bibitem{Chen:1999tn}
J.-W. Chen, G.~Rupak, and M.~J. Savage, ``{Nucleon-nucleon effective field
  theory without pions},'' {\em Nucl. Phys.}, vol.~A653, pp.~386--412, 1999,
  nucl-th/9902056.

\bibitem{Kong2000137}
X.~Kong and F.~Ravndal, ``Coulomb effects in low energy proton-proton
  scattering,'' {\em Nuclear Physics A}, vol.~665, no.~1-2, pp.~137 -- 163,
  2000.

\bibitem{Bedaque:1999ve}
P.~F. Bedaque, H.~W. Hammer, and U.~van Kolck, ``{Effective theory of the
  triton},'' {\em Nucl. Phys.}, vol.~A676, pp.~357--370, 2000, nucl-th/9906032.

\bibitem{Platter2005254}
L.~Platter, H.-W. Hammer, and U.-G. Mei{\ss}ner, ``On the correlation between
  the binding energies of the triton and the $\alpha$ particle,'' {\em Physics
  Letters B}, vol.~607, no.~3, pp.~254 -- 258, 2005.

\bibitem{Konig:2016utl}
S.~K{\"o}nig, H.~W. Grie{\ss}hammer, H.~W. Hammer, and U.~van Kolck, ``{Nuclear
  Physics Around the Unitarity Limit},'' {\em Phys. Rev. Lett.}, vol.~118,
  no.~20, p.~202501, 2017, 1607.04623.

\bibitem{Deltuva:2010xd}
A.~Deltuva, ``{Efimov physics in bosonic atom-trimer scattering},'' {\em Phys.
  Rev.}, vol.~A82, p.~040701, 2010, 1009.1295.

\bibitem{newton12}
R.~Newton, {\em Scattering Theory of Waves and Particles}.
\newblock Dover Publications, Inc., 2nd~ed., 2002.
\newblock Ch. 12.1.

\bibitem{PhysRevC.89.064003}
J.~Vanasse, D.~A. Egolf, J.~Kerin, S.~K{\"o}nig, and R.~P. Springer,
  ``$^{3}\mathrm{He}$ and $pd$ scattering to next-to-leading order in pionless
  effective field theory,'' {\em Phys. Rev. C}, vol.~89, p.~064003, Jun 2014.

\bibitem{Kirscher:2015zoa}
J.~Kirscher and D.~Gazit, ``{The Coulomb interaction in Helium-3: Interplay of
  strong short-range and weak long-range potentials},'' {\em Phys. Lett.},
  vol.~B755, pp.~253--260, 2016, 1510.00118.

\bibitem{Wiringa:2006ih}
R.~B. Wiringa, ``{Pair counting, pion-exchange forces, and the structure of
  light nuclei},'' {\em Phys. Rev.}, vol.~C73, p.~034317, 2006,
  nucl-th/0601064.

\bibitem{Stetcu:2006ey}
I.~Stetcu, B.~R. Barrett, and U.~van Kolck, ``{No-core shell model in an
  effective-field-theory framework},'' {\em Phys. Lett.}, vol.~B653,
  pp.~358--362, 2007, nucl-th/0609023.

\bibitem{newton17}
R.~Newton, {\em Scattering Theory of Waves and Particles}.
\newblock Dover Publications, Inc., 2nd~ed., 2002.
\newblock Ch. 17.2.2.

\bibitem{TILLEY19921}
D.~Tilley, H.~Weller, and G.~Hale, ``Energy levels of light nuclei a = 4,''
  {\em Nuclear Physics A}, vol.~541, no.~1, pp.~1 -- 104, 1992.

\bibitem{Wheeler:1937zza}
J.~A. Wheeler, ``{Molecular Viewpoints in Nuclear Structure},'' {\em Phys.
  Rev.}, vol.~52, pp.~1083--1106, 1937.

\bibitem{Wheeler:1937zz}
J.~A. Wheeler, ``{On the Mathematical Description of Light Nuclei by the Method
  of Resonating Group Structure},'' {\em Phys. Rev.}, vol.~52, pp.~1107--1122,
  1937.

\bibitem{Hofmann:1986}
H.~Hofmann in {\em Proceedings of Models and Methods in Few-Body Physics,
  Lisboa, Portugal} (L.~Ferreira, A.~Fonseca, and L.~Streit, eds.), p.~243,
  1986.

\bibitem{Kirscher:2015ana}
J.~Kirscher, ``{Pionless Effective Field Theory in Few-Nucleon Systems},''
  2015, 1506.00347.

\end{thebibliography}

\end{document}